\documentclass[a4paper,12pt]{article}

\usepackage[top=1.2in,right=0.5in,left=.5in,bottom=1.2in]{geometry}

\usepackage{latexsym}
\usepackage{amssymb}
\usepackage{float}
\usepackage{graphics,graphicx}
\usepackage{tabularx}
\usepackage{amsfonts}
\usepackage{amsmath}
\usepackage{enumerate}
\usepackage{booktabs}
\usepackage{multirow}
\usepackage{sectsty}
\usepackage[affil-it,auth-sc]{authblk}
\usepackage{subfig}
\usepackage{color}
\usepackage{mathtools}
\usepackage{mathrsfs}
\usepackage{bmpsize}
\usepackage{xcolor}
\usepackage{siunitx}
\usepackage{hyperref}

\sectionfont{\normalsize}
\subsectionfont{\normalsize}
\usepackage{fancyhdr}            
\fancypagestyle{plain}{%
  \fancyhead{}                                     
  \fancyfoot[CE,CO]{\thepage}       
  \fancyhead[CE,CO]{\textcolor[rgb]{0.64,0.15,0.15}{Published in \emph{Philosophical Transactions of the Royal Society A} (2022) 
  \\
doi: https://doi.org/10.1098/rsta.2022.0021}}
\setlength{\headheight}{14.5pt}}

\newcommand{\footmsg}[1]{%
  \let\temp\thempfn%
  \def\thempfs{}
  \footnotetext{#1}
  \let\tempfn\temp}

\begin{document}

\newcommand{\singlespace}{\baselineskip=12pt\lineskiplimit=0pt\lineskip=0pt}
\def\ds{\displaystyle}

\newcommand{\beq}{\begin{equation}}
\newcommand{\eeq}{\end{equation}}
\newcommand{\lb}{\label}
\newcommand{\ph}{\phantom}
\newcommand{\beqar}{\begin{eqnarray}}
\newcommand{\eeqar}{\end{eqnarray}}
\newcommand{\barr}{\begin{array}}
\newcommand{\earr}{\end{array}}
\newcommand{\jump}{\parallel}
\newcommand{\Ehat}{\hat{E}}
\newcommand{\That}{\hat{\bf T}}
\newcommand{\Ahat}{\hat{A}}
\newcommand{\chat}{\hat{c}}
\newcommand{\shat}{\hat{s}}
\newcommand{\khat}{\hat{k}}
\newcommand{\muhat}{\hat{\mu}}
\newcommand{\mc}{M^{\scriptscriptstyle C}}
\newcommand{\mei}{M^{\scriptscriptstyle M,EI}}
\newcommand{\mec}{M^{\scriptscriptstyle M,EC}}
\newcommand{\hbeta}{{\hat{\beta}}}
\newcommand{\rec}[2]{\left( #1 #2 \ds{\frac{1}{#1}}\right)}
\newcommand{\rep}[2]{\left( {#1}^2 #2 \ds{\frac{1}{{#1}^2}}\right)}
\newcommand{\derp}[2]{\ds{\frac {\partial #1}{\partial #2}}}
\newcommand{\derpn}[3]{\ds{\frac {\partial^{#3}#1}{\partial #2^{#3}}}}
\newcommand{\dert}[2]{\ds{\frac {d #1}{d #2}}}
\newcommand{\dertn}[3]{\ds{\frac {d^{#3} #1}{d #2^{#3}}}}
\newcommand{\ct}{\captionof{table}}
\newcommand{\cf}{\captionof{figure}}
\newcommand{\sgn}{\text{sgn}}

\def\c{{\circ}}
\def\bob{{\, \underline{\overline{\otimes}} \,}}
\def\ob{{\, \underline{\otimes} \,}}
\def\scalp{\mbox{\boldmath$\, \cdot \, $}}
\def\gdp{\makebox{\raisebox{-.215ex}{$\Box$}\hspace{-.778em}$\times$}}
\def\daa{\makebox{\raisebox{-.050ex}{$-$}\hspace{-.550em}$: ~$}}
\def\mK{\mbox{${\mathcal{K}}$}}
\def\cK{\mbox{${\mathbb {K}}$}}

\def\Xint#1{\mathchoice
   {\XXint\displaystyle\textstyle{#1}}%
   {\XXint\textstyle\scriptstyle{#1}}%
   {\XXint\scriptstyle\scriptscriptstyle{#1}}%
   {\XXint\scriptscriptstyle\scriptscriptstyle{#1}}%
   \!\int}
\def\XXint#1#2#3{{\setbox0=\hbox{$#1{#2#3}{\int}$}
     \vcenter{\hbox{$#2#3$}}\kern-.5\wd0}}
\def\ddashint{\Xint=}
\def\fpint{\Xint=}
\def\dashint{\Xint-}
\def\cpvint{\Xint-}
\def\intl{\int\limits}
\def\cpvintl{\cpvint\limits}
\def\fpintl{\fpint\limits}
\def\ointl{\oint\limits}
\def\bA{{\bf A}}
\def\ba{{\bf a}}
\def\bB{{\bf B}}
\def\bb{{\bf b}}
\def\bc{{\bf c}}
\def\bC{{\bf C}}
\def\bD{{\bf D}}
\def\bE{{\bf E}}
\def\be{{\bf e}}
\def\bbf{{\bf f}}
\def\bF{{\bf F}}
\def\bG{{\bf G}}
\def\bg{{\bf g}}
\def\bi{{\bf i}}
\def\bH{{\bf H}}
\def\bK{{\bf K}}
\def\bL{{\bf L}}
\def\bM{{\bf M}}
\def\bN{{\bf N}}
\def\bn{{\bf n}}
\def\bm{{\bf m}}
\def\b0{{\bf 0}}
\def\bo{{\bf o}}
\def\bX{{\bf X}}
\def\bx{{\bf x}}
\def\bP{{\bf P}}
\def\bp{{\bf p}}
\def\bQ{{\bf Q}}
\def\bq{{\bf q}}
\def\bR{{\bf R}}
\def\bS{{\bf S}}
\def\bs{{\bf s}}
\def\bT{{\bf T}}
\def\bt{{\bf t}}
\def\bU{{\bf U}}
\def\bu{{\bf u}}
\def\bv{{\bf v}}
\def\bw{{\bf w}}
\def\bW{{\bf W}}
\def\by{{\bf y}}
\def\bz{{\bf z}}
\def\T{{\bf T}}
\def\Te{\textrm{T}}
\def\Id{{\bf I}}
\def\bxi{\mbox{\boldmath${\xi}$}}
\def\balpha{\mbox{\boldmath${\alpha}$}}
\def\bbeta{\mbox{\boldmath${\beta}$}}
\def\bepsilon{\mbox{\boldmath${\epsilon}$}}
\def\bvarepsilon{\mbox{\boldmath${\varepsilon}$}}
\def\bomega{\mbox{\boldmath${\omega}$}}
\def\bphi{\mbox{\boldmath${\phi}$}}
\def\bsigma{\mbox{\boldmath${\sigma}$}}
\def\bfeta{\mbox{\boldmath${\eta}$}}
\def\bDelta{\mbox{\boldmath${\Delta}$}}
\def\btau{\mbox{\boldmath $\tau$}}
\def\tr{{\rm tr}}
\def\dev{{\rm dev}}
\def\div{{\rm div}}
\def\Div{{\rm Div}}
\def\Grad{{\rm Grad}}
\def\grad{{\rm grad}}
\def\Lin{{\rm Lin}}
\def\Sym{{\rm Sym}}
\def\Skw{{\rm Skew}}
\def\abs{{\rm abs}}
\def\Re{{\rm Re}}
\def\Im{{\rm Im}}
\def\capB{\mbox{\boldmath${\mathsf B}$}}
\def\capC{\mbox{\boldmath${\mathsf C}$}}
\def\capD{\mbox{\boldmath${\mathsf D}$}}
\def\capE{\mbox{\boldmath${\mathsf E}$}}
\def\capG{\mbox{\boldmath${\mathsf G}$}}
\def\tcapG{\tilde{\capG}}
\def\capH{\mbox{\boldmath${\mathsf H}$}}
\def\capK{\mbox{\boldmath${\mathsf K}$}}
\def\capL{\mbox{\boldmath${\mathsf L}$}}
\def\capM{\mbox{\boldmath${\mathsf M}$}}
\def\capR{\mbox{\boldmath${\mathsf R}$}}
\def\capW{\mbox{\boldmath${\mathsf W}$}}

\def\i{\mbox{${\mathrm i}$}}
\def\mC{\mbox{\boldmath${\mathcal C}$}}
\def\mB{\mbox{${\mathcal B}$}}
\def\mE{\mbox{${\mathcal{E}}$}}
\def\mL{\mbox{${\mathcal{L}}$}}
\def\mK{\mbox{${\mathcal{K}}$}}
\def\mV{\mbox{${\mathcal{V}}$}}
\def\C{\mbox{\boldmath${\mathcal C}$}}
\def\E{\mbox{\boldmath${\mathcal E}$}}

\def\AAM{{\it Advances in Applied Mechanics }}
\def\ACME{{\it Arch. Comput. Meth. Engng.}}
\def\ARMA{{\it Arch. Rat. Mech. Analysis}}
\def\AMR{{\it Appl. Mech. Rev.}}
\def\ASCEEM{{\it ASCE J. Eng. Mech.}}
\def\ACTA{{\it Acta Mater.}}
\def\CMAME {{\it Comput. Meth. Appl. Mech. Engrg.}}
\def\CRAS{{\it C. R. Acad. Sci. Paris}}
\def\CRM{{\it Comptes Rendus M\'ecanique}}
\def\EFM{{\it Eng. Fracture Mechanics}}
\def\EJMA{{\it Eur.~J.~Mechanics-A/Solids}}
\def\IJES{{\it Int. J. Eng. Sci.}}
\def\IJF{{\it Int. J. Fracture}}
\def\IJMS{{\it Int. J. Mech. Sci.}}
\def\IJNAMG{{\it Int. J. Numer. Anal. Meth. Geomech.}}
\def\IJP{{\it Int. J. Plasticity}}
\def\IJSS{{\it Int. J. Solids Structures}}
\def\IngA{{\it Ing. Archiv}}
\def\JAM{{\it J. Appl. Mech.}}
\def\JAP{{\it J. Appl. Phys.}}
\def\JAE{{\it J. Aerospace Eng.}}
\def\JE{{\it J. Elasticity}}
\def\JM{{\it J. de M\'ecanique}}
\def\JMPS{{\it J. Mech. Phys. Solids}}
\def\JSV{{\it J. Sound and Vibration}}
\def\MACRO{{\it Macromolecules}}
\def\MMT{{\it Mech. Mach. Th.}}
\def\MOM{{\it Mech. Materials}}
\def\MMS{{\it Math. Mech. Solids}}
\def\MMT{{\it Metall. Mater. Trans. A}}
\def\MPCPS{{\it Math. Proc. Camb. Phil. Soc.}}
\def\MSE{{\it Mater. Sci. Eng.}}
\def\NATURE{{\it Nature}}
\def\NATUREM{{\it Nature Mater.}}
\def\PHIL{{\it Phil. Trans. R. Soc.}}
\def\PMPS{{\it Proc. Math. Phys. Soc.}}
\def\PNAS{{\it Proc. Nat. Acad. Sci.}}
\def\PRE{{\it Phys. Rev. E}}
\def\PRL{{\it Phys. Rev. Letters}}
\def\PRSL{{\it Proc. R. Soc.}}
\def\RIIT{{\it Rozprawy Inzynierskie - Engineering Transactions}}
\def\ROCK{{\it Rock Mech. and Rock Eng.}}
\def\QAM{{\it Quart. Appl. Math.}}
\def\QJMAM{{\it Quart. J. Mech. Appl. Math.}}
\def\SCIENCE{{\it Science}}
\def\SCRMAT{{\it Scripta Mater.}}
\def\SM{{\it Scripta Metall.}}
\def\ZAMM{{\it Z. Angew. Math. Mech.}}
\def\ZAMP{{\it Z. Angew. Math. Phys.}}
\def\ZVDI{{\it Z. Verein. Deut. Ing.}}

\def\salto#1#2{
[\mbox{\hspace{-#1em}}[#2]\mbox{\hspace{-#1em}}]}

\renewcommand\Affilfont{\itshape}
\setlength{\affilsep}{1em}
\renewcommand\Authsep{, }
\renewcommand\Authand{ and }
\renewcommand\Authands{ and }
\setcounter{Maxaffil}{2}

\title{Buckling vs unilateral constraint\\
for a  multistable metamaterial element	}


\author{N. Hima$^{\text{A}, \text{B}}$, D. Bigoni$^{\text{A}}$ and F. Dal Corso$^{\textbf{A},}$}
 \affil[A]{DICAM, University of Trento, via~Mesiano~77, I-38123
Trento, Italy.}
 \affil[B]{FIP MEC srl, via~Scapacchi\`o~41, 35030 Selvazzano Dentro PD, Italy.}

\date{}
\maketitle \footnotetext[1]{Corresponding author: Francesco Dal Corso (francesco.dalcorso@unitn.it)}

\date{}
\maketitle

\begin{abstract}
A structural element is designed and investigated,  forming the basis for the development of an elastic multistable metamaterial. The leitmotif of the structural design is the implementation of a strut characterized by a bifurcation occurring at either vanishing tensile or compressive load. It is shown that  buckling at null load leads to a mechanical equivalence with a unilateral constraint formulation, introducing shocks in dynamics. Towards a future analysis of the latter,  the nonlinear quasi-static response is investigated, showing the multistable character of the structure, which may appear as bistable or tetrastable.
\end{abstract}


\section{Introduction}
Elastic metamaterials represent a blowing-up research field, finding crucial applications in vibration control, wave filtering and conditioning  
\cite{Guo2018, Bertoldi2017, Carta2016, Carta2018, DAlessandro2016, Adams2008, DAlessandro2019, Garau2018, Haslinger2017, Misseroni2019, Diatta2014, Cabras2017a, Babaee2016, Craster2013, DAlessandro2017, Milton2006, Misseroni2016, Willis2015}. 
However, meta-materials exhibit extraordinary mechanical properties even when subject to quasi-static loading, particularly when large deformations are involved 
\cite{Frenzel2017, Hou2014, Kadic2019, Kadic2019a, Liu2020, Lin2020, Ma2018, Zinco2020}. 
Examples are numerous, including cloaking \cite{Berger2017}, extreme stiffness \cite{Buckmann2014}, shape morphing \cite{Coulais2018}, auxeticity \cite{Mousanezhad2015, Zhang2020}, 
 negative thermal expansion \cite{Cabras2019}, and multistable architectures \cite{Shan, Singh, Khajehtourian, Medina, Yang2018}.
A flow in this research stream is the exploitation of structures beyond buckling and instability loads \cite{Reis2015}, in a range of extreme deformations 
\cite{Bigoni2015a}. Under these conditions, structures become \lq elastic machines', capable of realizing soft actuation \cite{Yang2015}, or developing propulsion forces 
\cite{Bigoni2014, TianChen}, 
or being used as soft devices \cite{Bosi2014}.

\begin{figure}[!h]
	\centering
	\includegraphics[width=0.9\textwidth]{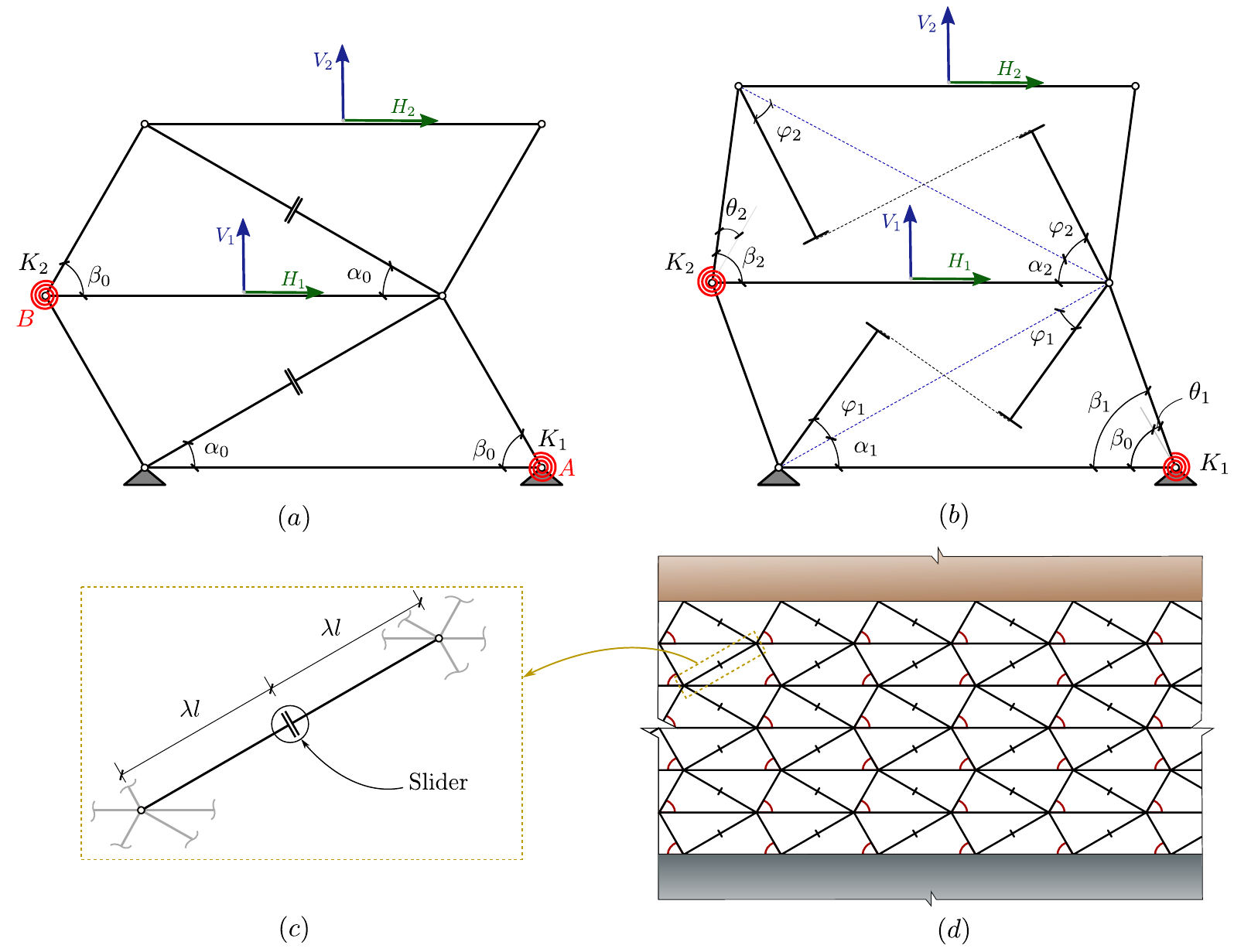}
	\caption{The unit structure of the proposed multistable  metamaterial element (sketched in the inset $(d)$). It is made up of two superimposed articulated quadrilateral structures, both composed of rigid bars and containing an elastic hinge (of stiffness $K_j$, $j=1,2$) and a slider at the midpoint of one of the inclined rigid bars. 
	The introduction of the slider  $(c)$ provides the key mechanical feature, because it buckles  at null axial force, realizing  a unilateral constraint in compression.
	($a$) Undeformed  and ($b$) deformed configuration of the two-layer structure subject to dead loads $H_j$ and $V_j$,  described through the misalignment angles  $\varphi_j$ or equivalently through the difference angles  $\theta_j$.}
	\label{fig:intro_sketch}
\end{figure}

The nonlinear analysis of the structure shown in Fig. \ref{fig:intro_sketch}($a$), forming an element to be exploited in a metamaterial design (Fig. \ref{fig:intro_sketch}($d$)), is the subject of the present article. The structure is composed of two superimposed layers of rigid bars  essentially working as 
quadrilateral linkages, but equipped with an elastic hinge and with a bar containing a slider (a constraint allowing only relative transverse displacement between the connected elements \cite{Zaccaria2011a}) which buckles under tensile load of vanishing magnitude. 

\begin{figure}[!h]
	\centering
	\includegraphics[width=\textwidth]{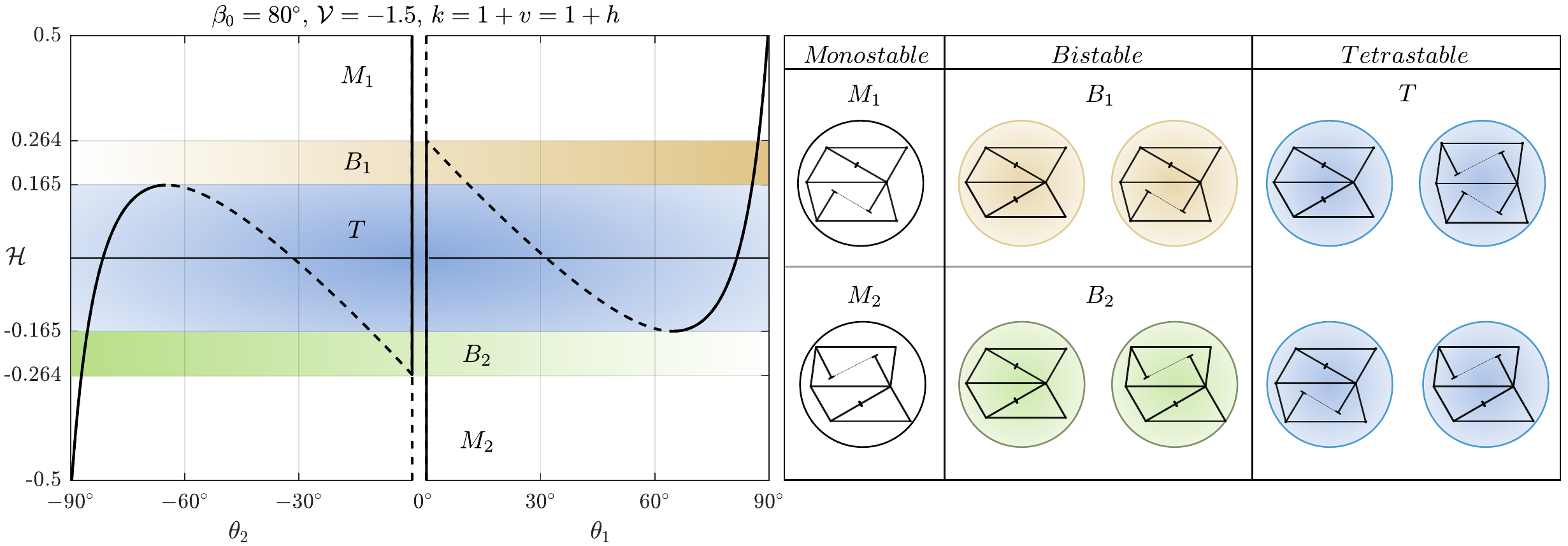}
	\caption{
	Equilibrium diagram for the structure shown in Fig. \ref{fig:intro_sketch}($a$) and ($b$), when subject to a constant vertical load $\mathcal{V}$, expressed in terms of the difference angles $\theta_1$ and $\theta_2$ as functions of the  variable horizontal load $\mathcal{H}$. Stable and unstable configurations are displayed as continuous and dashed curves, respectively. By varying the horizontal load $\mathcal{H}$, the system displays monostability ($M_1$ and $M_2$), bistability  ($B_1$ and $B_2$), and  tetrastability ($T$), through the corresponding  deformed configurations sketched beside.
	}
	\label{introcurves}
\end{figure}

The latter structural element 
provides the key mechanical feature implemented in the design of a simple structure, which displays a series of remarkable mechanical features, although characterized by only two degrees of freedom. 
These are related to the fact that the two structural layers can behave independently or synergically and involve a purely geometrical nonlinearity. In particular, the following features are found: (i.) the critical loads for  bifurcation depend on the geometry of the structure only through  the angle $\beta_0$, but are independent of the angle $\alpha_0$ and of the hinge stiffness $K_j$ ($j=1,2$); (ii.) the structure 
can have  multiple stable  equilibrium configurations under the same applied loads, which may be displayed as \emph{bistable} or \emph{tetrastable}  (as shown in Fig. \ref{introcurves}); (iii.) the applied loads can be varied in a way that a negative (a positive) slope in the load/displacement curve represents a stable (an unstable) loading path. 
Most of these features are related to the presence of an element  bifurcating in tension. However, it is shown 
that the a structure exhibiting exactly the same mechanical behaviour can be obtained through a proper modification of the quadrilateral linkages and substitution of the slider with a hinge, so that the inclined structural element suffers a bifurcation at a vanishing compressive load.
Interestingly,  the elements buckling at vanishing force can be replaced by unilateral constraints, which produce the same effect on the structure, but eliminates the bifurcation.
This  important aspect (to which the  next Section is dedicated) implies that the dynamic behaviour of the structure is characterized by the occurrence of impacts, a topic that will be analyzed elsewhere.

\section{Buckling vs unilateral constraint}\label{bif_or_unilateral}

\begin{figure}[!h]
	\centering
	\includegraphics[width=0.95\textwidth]{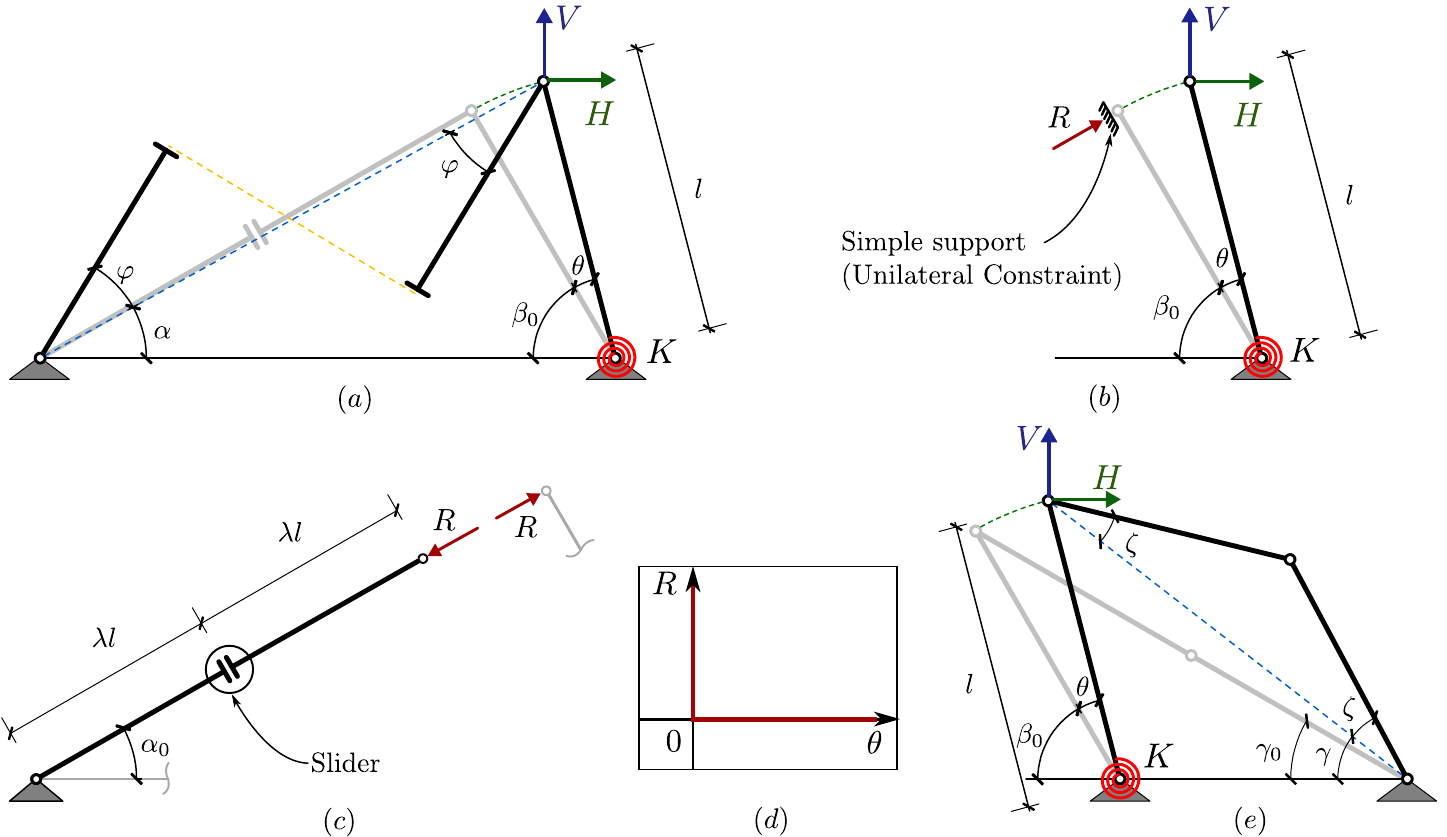}
	\caption{
	($a$) The essential building block (forming the two-layer unit structure shown in Fig. \ref{fig:intro_sketch}$a$) is 
	characterized by one elastic hinge of stiffness $K$ and is loaded with vertical $V$ and horizontal $H$ dead forces.
	The two structural elements, sketched in isolation in parts $(b)$ and $(c)$, are able to sustain an arbitrary amount of compression $R$, but they 
	cannot bear any tensile load, as one is equipped with a unilateral constraint and the other immediately buckles in tension (because it contains a slider). 
	Note that bifurcation does not play any role for unilateral contact ($b$), described by ($d$) the Signorini diagram.  ($e$) A quadrilateral linkage with two bars aligned parallel buckles in compression at null load, realizing an unilateral kinematics equivalent to the elementary structure ($a$).
    }
	\label{bifurcation_vs_unilateral}
\end{figure}

The elementary triangular structure reported in Fig. \ref{bifurcation_vs_unilateral}($a$) represents the essential 
building block 
of the two-layer unit structure depicted in 
Fig. \ref{fig:intro_sketch}($a$),
which in turn forms the 
structure leading to the 
interface shown in  Fig. \ref{fig:intro_sketch}$(d)$.
Although the rigid bar 
containing a slider at the mid-span is at equilibrium in its straight configuration  when axially loaded,  bifurcation occurs  at null axial force. 
Therefore, this single structural element displays an infinite stiffness under compression and 
a null stiffness under  tension. As a consequence, the mechanical behaviour of the rigid bar containing the 
slider becomes equivalent to a unilateral constraint, providing support only when compressed. This equivalence occurs only from the mechanical point of view, but not from a purely mathematical perspective. Indeed, the structure shown in Fig. \ref{bifurcation_vs_unilateral}($c$) is subject to a bifurcation, while that reported in Fig. \ref{bifurcation_vs_unilateral}($b$) is not. 
This crucial point is now explained in detail. 

For the structure in Fig. \ref{bifurcation_vs_unilateral}($a$), the total potential energy $\Pi$ can be written as the difference between the strain energy stored in the rotational spring of stiffness $K$ and the work done by the external dead loads ($H$ and $V$) 
\begin{equation}\label{TPE_theta}
    \Pi (\theta  ) = \dfrac{K \theta ^2 }{2} - H l \left[\cos \beta _0-\cos (\beta _0 + \theta )\right] - V l \left[\sin (\beta _0+\theta ) - \sin \beta _0 \right],
\end{equation}
where 
$l$ is the length of the  bar inclined at an angle $\beta_0$ in the undeformed configuration, with the latter subject 
to the following geometrical constraint
\begin{equation}\label{restangle}
 0<\beta_0<\pi,  
\end{equation}
while the difference angle $\theta$ is the Lagrangian parameter  subject to the unilateral constraint $\theta \geq 0$ and defining the inclination $\beta_0+\theta$ in the deformed configuration. Considering now the system in Fig. \ref{bifurcation_vs_unilateral}($b$), equivalent to the former one, the total potential energy $\Pi$ can be written as a function of another Lagrangian parameter, the misalignment angle $\varphi$  of the rigid bar with slider, as 
\begin{equation}
    \Pi (\varphi)= \Pi (\theta(\varphi)  ),
\end{equation}
where the difference angle $\theta$ is defined in relation of the misalignment angle $\varphi$ as
\begin{equation}\label{theta_phi}
        \theta(\varphi) = \arccos\left(\cos \beta _0 - \psi \tan ^2 \varphi \right)-\beta _0,
\end{equation}
a relation that cannot be inverted, as $\theta$ is insensitive to the sign of 
$\varphi$, $\theta(\varphi)=\theta(-\varphi)$, and with $\psi$  defined as 
\begin{equation}\label{psi}
        \psi = \frac{\sin \beta _0^2}{2 \sin \alpha_0  \sin (\alpha _0+\beta _0)}.
\end{equation}
Note that a Taylor series expansion of eq. (\ref{theta_phi})  about $\varphi=0$ truncated at the second-order leads to
\begin{equation}\label{fundamentalproperty}
\theta(\varphi)\approx \frac{\psi}{\sin \beta_0} \, \varphi^2, 
\end{equation}
highlighting the property that the two angle measures at small amplitude have different order. This last property implies that $\theta$ cannot be used as a Lagrangian parameter for the structure in Fig. \ref{fig:intro_sketch}($c$) and ($d$). 

Equilibrium of the structure sketched in Fig. \ref{bifurcation_vs_unilateral}($a$)  corresponds to the stationary condition for the total potential energy $\Pi(\varphi)$, which through the chain rule of differentiation becomes
\begin{equation}\label{Chain_Rule_phi}
\dfrac{\partial \Pi (\theta) }{\partial \theta } \dfrac{\partial \theta }{\partial \varphi } = 0,
\end{equation}
equivalent to 
\begin{equation}
\label{str}
\dfrac{\partial \Pi (\theta) }{\partial \theta }=0,\qquad \mbox{and/or}\qquad \dfrac{\partial \theta }{\partial \varphi } = 0. 
\end{equation}
Note that the second condition in eq. \eqref{str} 
becomes possible only when the order
of magnitude  in the relation between the parameters is different, as for the present structure as shown by \eqref{fundamentalproperty}.

From eq. (\ref{TPE_theta}) the condition  
\begin{equation}\label{Derivative_TPE_theta}
    \dfrac{\partial \Pi (\theta) }{\partial \theta } = K \theta - l \left[H \sin \left(\beta _0+\theta \right)+V \cos \left(\beta _0+\theta \right)\right],
\end{equation}
follows, while from eq. (\ref{theta_phi}) the derivative of $\theta$ with respect to $\phi$ can be written as
\begin{equation}\label{Derivative_Theta-phi}
    \dfrac{\partial \theta }{\partial \varphi } = 
    \dfrac{2\; \psi\tan \varphi }{\cos ^2\varphi \sqrt{ 1-\left(\cos \beta _0 - \psi  \tan ^2\varphi \right)^2} }.
\end{equation}
Expansions truncated at the second-order of the two above derivatives at small values of $\varphi$ provide
\begin{equation}\label{Derivative_TPE_theta_Expanded}
   \left. \dfrac{\partial \Pi (\theta) }{\partial \theta } \right|_{\theta(\varphi)}=  - l \left(H \sin \beta _0 +V \cos \beta _0\right),
   \qquad 
    \dfrac{\partial \theta }{\partial \varphi } = \dfrac{2 \; \psi }{\sin \beta _0} \varphi,
\end{equation}
showing that the equilibrium is not only attained for the trivial configuration for every load  combination
\begin{equation}
    \varphi=\theta=0\,\, \Rightarrow \,\,\dfrac{\partial \theta }{\partial \varphi } =0\,\,\Rightarrow \,\,
    \dfrac{\partial \Pi}{\partial \varphi } = 0\qquad\forall\,\, V \mbox{and}\,\, H,
\end{equation}
but  also for non-trivial configurations at the bifurcation condition
\begin{equation}\label{trivial_sol_in_phi}
H \sin\beta _0+V \cos \beta _0 = 0\,\, \Rightarrow \,\,\dfrac{\partial \Pi (\theta) }{\partial \theta } =0\,\,\Rightarrow \,\,
    \dfrac{\partial \Pi}{\partial \varphi } = 0\qquad\forall\,\, \varphi\neq 0\,\, (\mbox{with}\,\, \varphi^2\ll\varphi).
\end{equation}
Stability of the trivial configuration can be analyzed by considering the sign of the second derivative of $\Pi(\varphi)$, calculated at $\varphi=0$ as
\begin{equation}
\label{peretta}
    \left.\frac{\partial^2 \Pi(\varphi)}{\partial \varphi^2}\right|_{\varphi = 0} = -
    \frac{l \sin \beta_0}{\sin \alpha_0 \sin(\alpha_0 + \beta_0)}
     (H \sin \beta_0+V\cos \beta_0).
\end{equation}
Considering that 
\begin{equation}\label{restangle_2}
\alpha_0>0,\qquad
\alpha_0+\beta_0<\pi,
\end{equation}
eq. \eqref{peretta} shows that  the trivial configuration is stable whenever 
\begin{equation}
\label{seghina}
H \sin \beta_0+V\cos \beta_0 <0,
\end{equation}
and otherwise is unstable. Therefore, the structural element containing the slider, Fig. \ref{bifurcation_vs_unilateral}($c$),  essentially works as that subject to the unilateral constraint, Fig. \ref{bifurcation_vs_unilateral}($b$), because it buckles at null axial force and does not bear any tensile load. 

The equilibrium equation for the structure subject to the unilateral constraint,  Fig. \ref{bifurcation_vs_unilateral}($b$), is given by
\begin{equation}
\label{seghetta}
\dfrac{\partial \Pi (\theta) }{\partial \theta } \, \delta \theta \geq 0, 
\end{equation}
for all virtual displacements $\delta \theta \geq 0$, so that the unilateral Kuhn-Tucker conditions  are satisfied
\begin{equation}
\theta \geq 0, ~~~ R \geq 0, ~~~ \theta \, R =0,
\end{equation}
where $R$ is the reaction of the unilateral support. 
Equation \eqref{seghetta} holds with the \lq $=$' sign for all $\theta >0$ and 
becomes equivalent to eq. \eqref{seghina}, except that 
\lq $<$' has to be replaced with \lq $=$'.  
Therefore, {\it the presence of the unilateral constraint eliminates the bifurcation}, so that the buckling analysis is turned into a purely equilibrium  problem.  
The unilateral constraint is smooth, so that it does not alter the conservativeness of the system (which is subject to dead loading in the present formulation). For this reason, the Dirichlet stability theorem applies even in the boundary case $\theta=0$. There, the total potential energy is allowed to possess a non-analytical minimum to verify  stability. Therefore, positiveness of the first derivative of eq. \eqref{TPE_theta}, eq. \eqref{Derivative_TPE_theta} evaluated at $\theta=0$, leads  exactly to the 
same condition as in eq. \eqref{seghina}, thus confirming that both structures shown in Fig. \ref{bifurcation_vs_unilateral}($a$) and ($b$) exhibit the same mechanical behaviour. 
Note that for simplicity, the term \emph{bifurcation} will be associated in the following to both the misalignment and difference angles, $\varphi$ and $\theta$.

\paragraph{Buckling in tension vs compression.}
The kinematics of the inclined bar of length $l$ described by $\theta>0$ has been considered as the result of tensile buckling  in the simple triangular structure in Fig. \ref{bifurcation_vs_unilateral}($a$). The same kinematics for the inclined bar of length $l$ can be, however, equivalently achieved  through buckling in compression as for the  elementary triangular structure in Fig. \ref{bifurcation_vs_unilateral}($e$), now  incorporating a bar with a hinge (instead of a slider), initially straight and inclined at $\gamma_0$ in the undeformed state. 
In this latter structure, the difference angle $\theta$ becomes the following function of the  misalignment angle $\zeta$ 
\begin{equation}
    \theta(\zeta)= \arccos\left[ \dfrac{\left(\cos ^2\zeta - \cos ^2\gamma_0 \right) \sin ^2\beta _0}{2 \sin \gamma_0 \sin \left(\beta _0-\gamma _0\right)} + \dfrac{\sin \left(2 \beta _0\right) \cos \gamma _0-\left(\cos ^2 \beta_0 + 1\right) \sin \gamma _0}{2 \sin \left(\beta _0-\gamma _0\right)}
    \right]-\beta _0,
\end{equation} 
whose expansion for small    values of $\zeta$ simplifies to
\begin{equation}
\theta(\zeta)\approx \dfrac{\sin \beta _0}{2 \sin \gamma _0 \sin \left(\beta _0-\gamma _0\right)}  \zeta^2,
\end{equation}
showing the different order in the difference  $\theta$ and misalignment $\zeta$ angles,  in analogy to the misalignment angle $\varphi$ in the  triangular structure with tensile buckling, eq. (\ref{fundamentalproperty}).

Since the total potential energy $\Pi(\theta)$ (\ref{TPE_theta}) is the same for the two triangular structures 
shown in Figs. \ref{bifurcation_vs_unilateral}($a$) and ($e$)
when the  inclined bar is subject to the same rotation $\theta(\varphi)=\theta(\zeta)$,
the two structures,  although  based on two different types of buckling, are mechanically equivalent. For this reason, the results presented in the following and obtained for the system composed of layers with inclined elements displaying tensile buckling also holds for the analogous structure  whose deformation is linked to compressive buckling.

\section{Mechanics of the structure}

The mechanics of the planar structure sketched in Fig. \ref{fig:intro_sketch}$(a)$ and $(b)$ is investigated. The system combines two superimposed articulated quadrilateral structures, made up of rigid bars connected to each other through hinges in a parallelogram shape, and equipped with a slider (imposing continuity of rotational and axial displacement, but allowing a jump in the transverse displacement) at the mid-span on the diagonal bars.
The slider inside the $j$-th layer (two layers are considered, so that $j = 1, 2$) is activated when the related misalignment angle $\varphi_j$ assumes a non-null value. 
After bifurcation, the kinematics of the $j$-th layer, maintaining the shape of a parallelogram (the horizontal bars are subject to pure translational motion), is described by the two angles $\alpha_j$ and $\beta_j$ (configuration angles), both functions of $\varphi_j$ as 
\begin{equation}\label{gen_conf}
\alpha_j(\varphi_j) = \arccos\left(\dfrac{ \psi}{2 \lambda \cos \varphi_j} + \chi \cos \varphi_j\right),\qquad\beta_j(\varphi_j) = \arccos\left(\cos \beta _0 - \psi \tan ^2\varphi_j \right),
\end{equation}
where $\lambda$ and $\chi$ are constants depending on the initial configuration angles as follows 
\begin{equation}\label{lambda_chi}
    \lambda = \dfrac{\sin \beta_0}{2 \sin \alpha_0}>0, \qquad \chi = \dfrac{ \sin (2\alpha_0 + \beta_0)}{\sin (\alpha_0 + \beta_0)},
\end{equation}
while $\psi$ is as defined in eq. \eqref{psi}. Note that $\alpha_0$ and $\beta_0$ correspond to $\alpha_j$ and $\beta_j$ measured in the undeformed configuration described by  $\varphi_j=0$ and are subject to the geometrical constraints defined in eqs. \eqref{restangle} and \eqref{restangle_2}.

The deformed state is subject to restoring forces provided by the linear elastic rotational springs of stiffness $K_1$ and $K_2$, located at points indicated with letters \lq A' and \lq B', and unloaded in the undeformed configuration. 
The $j$-th layer is subject to the  horizontal $H_j$ and vertical $V_j$ dead loads, acting at the middle of the horizontal upper bar. To simplify the presentation, the following dimensionless  loads and the rotational stiffness ratios are introduced
 \begin{equation}\label{load_ratios}
	\begin{array}{cc}
		h=\dfrac{H_1}{H_2}, \qquad v = \dfrac{V_1}{V_2}, \qquad k = \dfrac{K_1}{K_2}\geq 0.
	\end{array}
\end{equation}
From the above description, it follows that the system configuration is entirely described by the evolution of two degrees of freedom, namely, the misalignment angles $\varphi_j$ ($j =1, 2$). Due to the properties of the system and its similarities with a structure with a unilateral constraint shown in the previous Section, it is expedient to make reference to the  difference angles $\theta_j=\beta_j-\beta_0$,  evaluated for both layers similarly to eq. \eqref{theta_phi} as
\begin{equation}\label{thetaj_phi}
        \theta_j(\varphi_j) =-(-1)^{j} \left[ \arccos\left(\cos \beta _0 - \psi \tan ^2 \varphi_j\right)-\beta _0\right],
\end{equation}
which yields the following constraints
 \begin{equation}\label{dbeta_conditions}
 	\theta_1\geq 0,\qquad
 	\theta_2\leq 0,\qquad\mbox{which imply}\qquad
 	\theta_1\theta_2\leq 0.
 \end{equation}
Finally, it is worth to highlight that the special case of the $j$-th parallelogram reducing to a line segment is provided by  the following condition for the difference angle
\begin{equation}
\left|\theta_j\right|=\overline{\theta}^{[n]},\qquad
\mbox{where}\qquad
\overline{\theta}^{[n]}=n\pi-\beta_0,\qquad n\in\mathbb{N},
\end{equation}
implying that the misalignment angle $\varphi_j$ is bounded as
\begin{equation}\label{max_disp}
		\left|\varphi_j\right|\leq \overline{\varphi},\qquad
\mbox{where}\qquad
		\overline{\varphi} = \arctan \sqrt{\dfrac{1+\cos\beta_0 }{\psi}}.
\end{equation}

\section{Total potential energy and equilibrium}

With reference to the misalignment angle $\varphi_j$, the total potential energy $\Pi$ of the system sketched in panels $(a)$ and $(d)$ of Fig. \ref{fig:intro_sketch}, is given as the summation of the elastic energy stored in the two elastic hinges and the negative of the work done by the forces acting on each layer
\begin{equation}\label{TOTpot_eng}
	\begin{array}{lll}
		\Pi(\varphi_1, \varphi_2) = \displaystyle K_2 \sum_{j=1}^{2}&\displaystyle\left\{\dfrac{k^{2-j}\left[\theta_j(\varphi_j)\right]^2}{2}  +  (1+v)^{2-j} \mathcal{V}  \left[ \sin \beta _0 -\sin \left(\beta_0 -(-1)^j\theta_j(\varphi_j)\right) \right]\right.\\[2mm]
		& \left.  + (-1)^{j}  (1+h)^{2-j} \mathcal{H} \left[\cos \beta _0-\cos \left(\beta _0 -(-1)^j \theta_j(\varphi_j) \right)\right] \right\},
	\end{array}
\end{equation}
where $k$, $h$ and $v$ are the stiffness and loading ratios  defined in eq. \eqref{load_ratios}, while $\mathcal{H}$ and $\mathcal{V}$ are the dimensionless horizontal and vertical dead loads acting on the upper layer, defined as
\begin{equation}\label{dim_less}
	\begin{array}{cc}
		\mathcal{H}=\dfrac{H_2 l}{K_2}, \qquad \mathcal{V}=\dfrac{V_2 l}{K_2}.
	\end{array}
\end{equation}
Further, the system of equilibrium equations can be obtained through the vanishing of the gradient of $\Pi$, eq. \eqref{TOTpot_eng}, with respect to the misalignment angles $\varphi_j$, 
\begin{equation}\label{full_chain_phi_theta0}
    \begin{array}{cc}
        \dfrac{\partial \Pi (\varphi_1, \varphi_2) }{\partial \varphi_j }=0,
    \end{array}
\end{equation}
which through  the chain rule of differentiation simplifies as
\begin{equation}\label{full_chain_phi_theta}
    \begin{array}{cc}
       \dfrac{\partial \Pi (\theta_1, \theta_2) }{\partial \theta_1}  \dfrac{\partial \theta_1}{\partial \varphi_j}+\dfrac{\partial \Pi (\theta_1, \theta_2) }{\partial \theta_2}  \dfrac{\partial \theta_2}{\partial \varphi_j}=0.
    \end{array}
    \end{equation}
Considering eq. \eqref{thetaj_phi} the following property holds for the structure
\begin{equation}\label{no_coupling}
    \dfrac{\partial\theta_q}{\partial\varphi_j} = \left\{
    \begin{array}{lll}
        0, & q \neq j,\\[2mm]
        -(-1)^j \;    \dfrac{2\; \psi\tan \varphi_j }{\cos ^2\varphi_j \sqrt{ 1-\left(\cos \beta _0 - \psi  \tan ^2\varphi_j\right)^2} }, & q = j,
    \end{array}\right.
\end{equation}
and therefore the equilibrium equation \eqref{full_chain_phi_theta0} is reduced to
\begin{equation}\label{chanin_phi_theta}
    \begin{array}{cc}
        \dfrac{\partial \Pi (\theta_1, \theta_2) }{\partial \theta_j}  \dfrac{\partial \theta_j}{\partial \varphi_j}=0,
    \end{array}
\end{equation}
where a repeated index does not imply summation, here and henceforth. The equilibrium equations (\ref{chanin_phi_theta}) are always satisfied for the trivial configuration $\varphi_j=\theta_j=0$ because
\begin{equation}
\left.\dfrac{\partial \theta_j}{\partial \varphi_j}\right|_{\varphi_j=0}=0.
\end{equation}
For non-trivial configurations, the equilibrium equations (\ref{chanin_phi_theta})  in terms of the misalignment angle $\varphi_j$ are quite complex, and therefore are not reported because impractical. However,  making use of the equivalence between the two elementary structures reported in Fig. \ref{bifurcation_vs_unilateral}, it is expedient to write the equilibrium conditions of the structure in a non-trivial configuration by utilizing as a parameter the difference angle $\theta_j\neq 0$ through
\begin{equation}
\dfrac{\partial \Pi (\theta_1, \theta_2) }{\partial \theta_j} =0,
\end{equation} 
providing
\begin{equation}\label{static_eqn_theta}
        \begin{array}{lll}
    	   (h+1) \mathcal{H} \sin(\beta_0+\theta_1) + (v+1)  \mathcal{V} \cos \left(\beta _0+\theta _1 \right) = k \; \theta _1, \quad &\mbox{for} \; \theta_1 > 0, \\[3mm]
    	   \mathcal{H} \sin \left(\beta _0 - \theta _2\right) - \mathcal{V} \cos \left(\beta _0 - \theta _2\right) = \theta _2, \quad & \mbox{for} \; \theta_2 < 0.
    	\end{array}
\end{equation}

Similarly to the simple case explained in Section \ref{bif_or_unilateral}, the difference angles $\theta_j$ do not represent Lagrangian parameters for the two-layer unit structure shown in Fig. \ref{fig:intro_sketch} and therefore $\theta_j=0$ is not a solution for the system of equations \eqref{static_eqn_theta}.
In terms of the unilateral constraint model, when one or both of the layers composing the planar structure remain undeformed, a reaction emerges preventing the mechanism to move towards the constrained direction, but does not pose any obstacle to the opposite movement ($\theta_1 \geq 0$ and $\theta_2\leq0$). The unilateral constraints impose the following Kuhn-Tucker conditions          
\begin{equation}\label{Kuhn-Tucker}
    \begin{array}{cc}
        \theta_1 \geq 0, \qquad \mathcal{R}_1 \geq 0, \qquad \theta_1 \mathcal{R}_1=0, \\ [5mm]
        \theta_2 \leq 0, \qquad \mathcal{R}_2 \leq 0, \qquad \theta_2 \mathcal{R}_2=0, 
    \end{array}
\end{equation}
where $\mathcal{R}_1$ and $\mathcal{R}_2$ are the reactions emerging in the constraint present within the respective layers, and are equivalent to the compression force in the bars with the slider.    
It is also interesting to note that  equilibrium equations (\ref{static_eqn_theta}) reveal that the  2 degrees of freedom of the structure are decoupled. This implies that \emph{the equilibrium of the two layers  relies on the independent \lq individual' equilibrium of each layer}. Therefore,  depending on the existence of the non-trivial configuration for each layer, the  equilibrium configurations for the entire structure can be distinguished in:
\vspace{-4mm}
\begin{itemize}
    \item  trivial configuration 
    ($\theta_1=\theta_2= 0$, always existing);
        \item   non-trivial configuration for only the lower ($\theta_1= 0$ and $\theta_2< 0$) or the upper  ($\theta_1> 0$ and $\theta_2= 0$) layer;
            \item  non-trivial configuration for  both layers ($\theta_1> 0$ and $\theta_2< 0$).
\end{itemize}
\vspace{-4mm}
Non-trivial configurations are investigated in the next Section, along with the bifurcation conditions and the stability of  equilibrium.

\section{Bifurcation, equilibrium, and multistability}

Bifurcation, post-critical behaviour, and stability of the two-layer unit structure shown in Fig. \ref{fig:intro_sketch} are analyzed at varying dimensionless  horizontal $\mathcal{H}$ and vertical $\mathcal{V}$ dead loads. 

\subsection{Critical loads and post-critical response}
As mentioned, the equilibrium equations \eqref{full_chain_phi_theta} are satisfied by the trivial  configuration ($\varphi_j=\theta_j=0$), while  for the non-trivial configuration ($\varphi_j \neq 0$) reference can be made to the equilibrium condition expressed in terms of $\theta_j$, eq.  \eqref{static_eqn_theta}. 

The linearized version of the equilibrium equations \eqref{static_eqn_theta} can be obtained through a Taylor series expansion about $\theta_j=0$ as 
\begin{equation}\label{eq:eq_lin_dual_beta}
    	\begin{array}{llll}
    	    &(h+1) \mathcal{H} \sin \beta_0 + (v+1)\mathcal{V} \cos \beta_0 = \left(k + \dfrac{(v+1) \mathcal{V}}{\sin \beta_0} \right) \theta_1, \qquad &\mbox{for} \; \theta_1>0,\\[4mm]
    	    &\mathcal{H} \sin \beta_0 - \mathcal{V}\cos \beta_0 = \left(1 + \dfrac{\mathcal{V}}{\sin \beta_0} \right) \theta_2,\qquad &\mbox{for} \; \theta_2<0.
    	\end{array}
\end{equation}
The load pairs $\mathcal{H}$ and $\mathcal{V}$, solution of eq. \eqref{eq:eq_lin_dual_beta} at vanishing $\theta_j$, represent the set of bifurcation conditions for the $j$-th layer, namely, the \emph{critical load combinations}. The critical pair of loads corresponds to 
\begin{equation}
\label{asymetric_act_load}
	 \left\{
	\begin{array}{cc}
	h+1 \\[2mm]
	1	\end{array}
	\right\}\mathcal{H}_{cr} \sin \beta_0
	+ 
	 \left\{
	\begin{array}{cc}
	v+1 \\[2mm]
	-1	\end{array}
	\right\}\mathcal{V}_{cr}\cos \beta_0  = 0,
	~~~
	\begin{array}{lll}
	\mbox{for lower layer bifurcation}, \\[2mm]
	\mbox{for upper layer bifurcation},	\end{array}
\end{equation}
showing independence of the stiffness of the rotational springs, so that {\it the system bifurcation  is purely geometrical and only involves the referencial configuration  angle $\beta_0$}. 

From eq. (\ref{asymetric_act_load}), 
 the  following skew-symmetric behaviour is displayed 
\begin{equation}\label{act_load_mirror}
	\begin{array}{lll}
		\mathcal{H}_{cr}(\mathcal{V}_{cr},\beta_0) = -\mathcal{H}_{cr}(\mathcal{V}_{cr},\pi-\beta_0),
	\end{array}
\end{equation}
and the simultaneous bifurcation of  both layers occurs when 
\begin{equation}\label{bif_nobif}
		h+v = -2.
\end{equation}

\begin{figure}[!h]
    \centering
	\includegraphics[width=.95\textwidth]{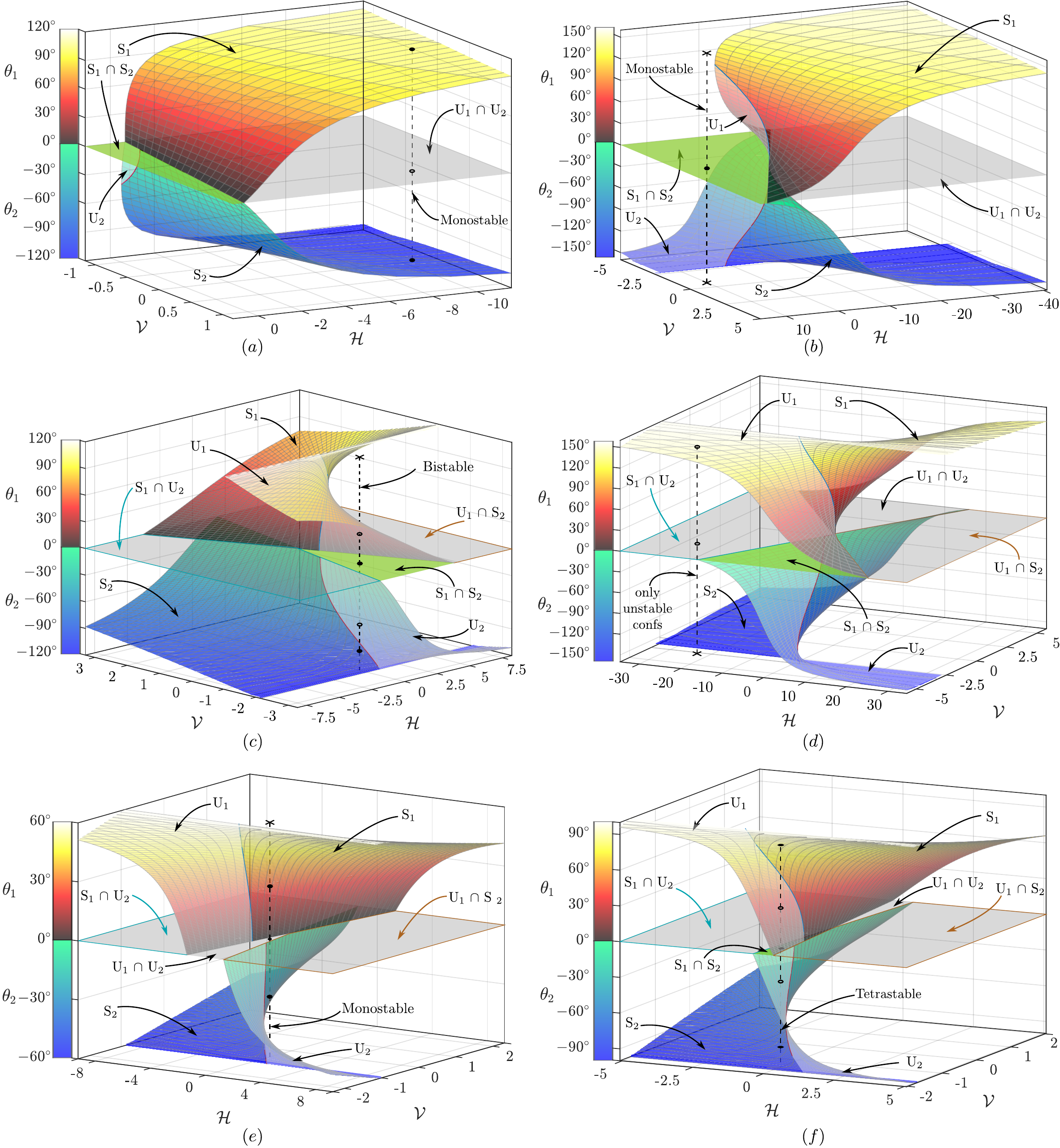}
        \caption{Equilibrium configurations in the  $\mathcal{H}$--$\mathcal{V}$--$\theta_j$ space for the system described by the parameters sets ($a$)--($f$) reported in Table \ref{tab:cases}. Stable and unstable configurations for the $j$-th layer are idenitified through the letters \lq U$_j$' or \lq S$_j$'.
        Bifurcation loads $\mathcal{H}_{cr}$--$\mathcal{V}_{cr}$  are given by the intersection of the non-trivial configuration surfaces with the trivial configuration plane ($\theta_j=0$). Uniqueness, non-uniqueness, absence of stable configurations, monostability, bistability  and tetrastability (within the limited  range $|\theta_j|<\pi-\beta_0$ and) associated to specific loads combinations $\mathcal{H}$--$\mathcal{V}$ (vertical dashed lines) are highlighted.}
        \label{fig:eq_surf}
\end{figure}

The post-buckling behaviour in terms of the difference angles $\theta_1$ and $\theta_2$ as functions of $\mathcal{V}$ and $\mathcal{H}$, obtained as the  solution of the nonlinear   eqs. \eqref{static_eqn_theta},
is depicted in Fig. \ref{fig:eq_surf} (in the limited  range $|\theta_j|<\pi-\beta_0$)  for the six  sets of parameters reported in Table \ref{tab:cases}, the first four associated to a non-symmetric  response, while the last two correspond to a symmetric one.

    \begin{table}[!h]
    \caption{Sets of structural parameters corresponding to the equilibrium configurations reported in Fig. \ref{fig:eq_surf}.}
    \label{tab:cases}
    \begin{center}
    \begin{tabular}{lllll}
    \hline
    Panel  &  $\beta_0$ & $h$ & $v$ & $k$ \\
    \hline
    ($a$) & 60$^\circ$ & -2.5 & 0.5 & 2 \\
    ($b$) & 20$^\circ$ & -2 & 0 & 3\\
    ($c$) & 60$^\circ$ & 0 & 1 & 1 \\
    \hline
    \end{tabular}\qquad
    \begin{tabular}{lllll}
    \hline
     Panel &  $\beta_0$ & $h$ & $v$ & $k$\\
    \hline
    ($d$) & 20$^\circ$ & 1.5 & 3 & 3\\
    ($e$) & 120$^\circ$ & 0 & 0 & 1\\
    ($f$) & 80$^\circ$ & 0 & 0 & 1 \\
    \hline
    \end{tabular}
    \end{center}
\end{table}

A total of three equilibrium surfaces is shown: the trivial equilibrium plane ($\theta_1=\theta_2=0$) and two equilibrium surfaces corresponding to the non-trivial configuration for the lower ($\theta_1>0$) and the upper layer ($\theta_2<0$). The stability character (addressed in Section \ref{stability})  of the portions of these equilibrium surfaces is also indicated with the letter \lq U$_j$' or \lq S$_j$', respectively denoting the \lq unstable' or \lq stable' configuration for the $j$-th layer (namely, U$_1$ defines an unstable configuration for the lower layer).

 Fig. \ref{fig:eq_surf} can be interpreted in the following way: (stable or unstable) equilibrium configurations for $\theta_1$ and $\theta_2$  correspond to the intersections of the equilibrium surfaces with the (vertical dashed) line, defined by constant values of  $\mathcal{H}$ and $\mathcal{V}$, representing a load combination applied to the structure. Every vertical dashed line always intersects  the trivial equilibrium surface, while depending on the structural and loading parameters it may  intersect none, one, or  multiple times (in the considered difference angle range) the non-trivial surfaces.

Thus, with reference to the systems reported in  Fig. \ref{fig:eq_surf}, examples of loads combination show different numbers of intersections and consequently of equilibrium states. More specifically:
\begin{itemize}
    \item  for the load combination highlighted in Fig. \ref{fig:eq_surf}($a$) and ($e$), the vertical dashed line intersects 3 times the equilibrium surfaces, which correspond to four possible equilibrium configurations: (i.) $\theta_1 > 0$ and $\theta_2 < 0$, (ii.) $\theta_1 = 0$ and $\theta_2 < 0$,  (iii.) $\theta_1 > 0$ and $\theta_2 = 0$ and finally (iv.) $\theta_1 = \theta_2=0$. Among these, only (i.) is stable, which corresponds to a deformation involving both layers;
    \item  for the load combination highlighted in Fig. \ref{fig:eq_surf}($b$), the vertical dashed line intersects only the trivial equilibrium surface, corresponding to the stable trivial configuration for both layers;
    \item  for the load combination highlighted in Fig. \ref{fig:eq_surf}($c$), the vertical dashed line intersects 4 times the equilibrium surfaces, 
    leading to 6 equilibrium configurations: (i.) two sets of $\theta_1 > 0$ and $\theta_2 < 0$, (ii.) two sets of $\theta_1 = 0$ and $\theta_2 < 0$,  (iii.) $\theta_1 > 0$ and $\theta_2 = 0$, and (iv.) the trivial state $\theta_1 = \theta_2=0$. Among these, only 2 configurations are stable: one of the sets in (ii.) and the trivial configuration  (iv.), so that the system is \emph{bistable} under the highlighted $\mathcal{H}$--$\mathcal{V}$ loads combination;
    \item  for the load combination highlighted in Fig. \ref{fig:eq_surf}($d$), the vertical dashed line intersects 2 times  the equilibrium surfaces  and  2 equilibrium configurations exist: (i.)  $\theta_1> 0$ and $\theta_2 = 0$ and (ii.) $\theta_1 = \theta_2 = 0$. Among these, none is stable;
    \item  for the load combination highlighted in Fig. \ref{fig:eq_surf}($f$), the vertical dashed line intersects 5 times the equilibrium surfaces, providing a total of 9 different equilibrium configurations: (i.)  four sets of $\theta_1 > 0$ and $\theta_2 < 0$, (ii.) two sets of $\theta_1 = 0$ and $\theta_2 < 0$,  (iii.) two sets of $\theta_1 > 0$ and $\theta_2 = 0$ and (iv.) the trivial state $\theta_1 = \theta_2=0$. Among these, 4 configurations are stable:  one per each of the sets (i.), (ii.), (iii.) and the trivial configuration (iv.), so that the system is \emph{tetrastable} under the highlighted $\mathcal{H}$--$\mathcal{V}$ load  combination.
\end{itemize}

\begin{figure}[!h]
	\centering
	\includegraphics[width=\textwidth]{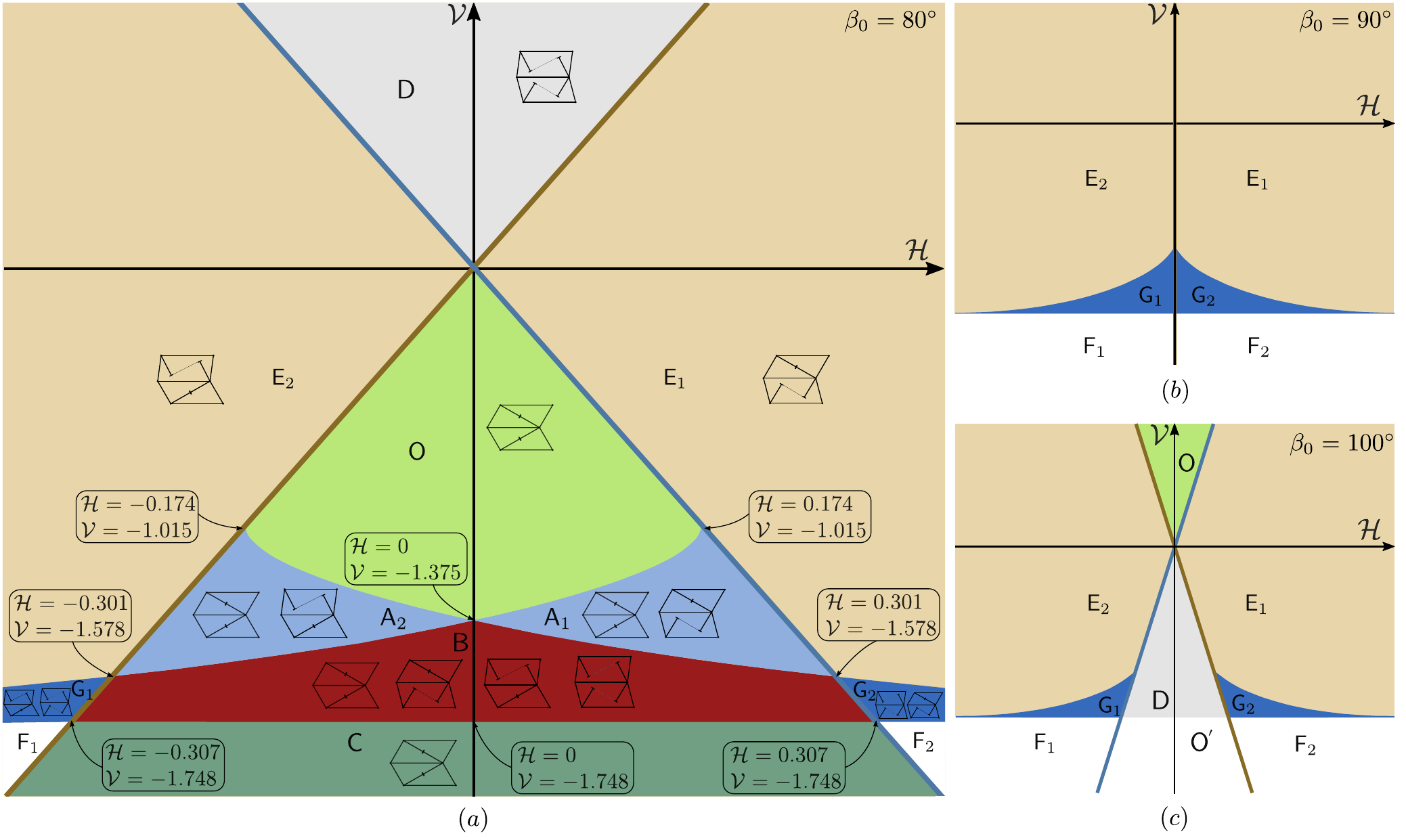}
		\caption{($a$) Projection onto the  $\mathcal{H}-\mathcal{V}$ plane of the equilibrium surfaces (limited to $|\theta_j|<\pi-\beta_0$) reported in Fig. \ref{fig:eq_surf}($f$), corresponding to  the parameters set ($f$) of Table \ref{tab:cases}.  
		Stable equilibrium configurations are sketched inside each region, except for regions $\mathsf{F}_j$, where stable configuration do not exist. Regions with the same letter, but differing because the $j$-th layer displays a non-trivial configuration, are distinguished through the subscript $j$. Panels ($b$) and ($c$) are as panels ($a$), except that $\beta_0=90^\circ$ and  $\beta_0=100^\circ$, respectively. The number and  properties of the equilibrium configurations corresponding to the different nine regions are listed in Table \ref{tab:regions}.}
	\label{fig:bifurcation_regions}
\end{figure}

The non-uniqueness of equilibrium configurations (restricted  to $|\theta_j|<\pi-\beta_0$) reported in Fig. \ref{fig:eq_surf}($f$) can be further appreciated through their projection onto the  $\mathcal{H}$--$\mathcal{V}$ plane, as reported in Fig. \ref{fig:bifurcation_regions}($a$). The influence of the angle $\beta_0$  is shown through   the  complementary projections   corresponding to $\beta_0=90^\circ$, Fig. \ref{fig:bifurcation_regions}($b$), and to $\beta_0=100^\circ$, Fig.  \ref{fig:bifurcation_regions}($c$), 
with the other structural and loading parameters remaining the same as in Fig. \ref{fig:bifurcation_regions}($a$). 

A total of nine regions for $\mathcal{H}$--$\mathcal{V}$ load combinations  are distinguished,  corresponding to a different number and type of  equilibrium configurations. The corresponding stable configurations are sketched  for each region visible in Fig. \ref{fig:bifurcation_regions}($a$).  The properties of the equilibrium configurations of these nine regions are summarized in Table \ref{tab:regions},
showing that, when the load combination $\mathcal{H}$--$\mathcal{V}$ varies, under the restriction  $|\theta_j|<\pi-\beta_0$, the system changes the number of its stable equilibrium configurations and in particular can display: none (regions $\mathsf{O}'$ and $\mathsf{F}$), one (\emph{monostable}, regions $\mathsf{O}$, $\mathsf{C}$, $\mathsf{D}$, and $\mathsf{E}$), two (\emph{bistable}, regions $\mathsf{A}$ and $\mathsf{G}$), or four (\emph{tetrastable}, region $\mathsf{B}$)  stable equilibrium configurations.

 \begin{table}[!h]
 \caption{\label{tab:regions} Number of the equilibrium configurations and their stability corresponding to the regions of load combination $\mathcal{H}$--$\mathcal{V}$ reported in Fig. \ref{fig:bifurcation_regions}.}
 \begin{center}
 \begin{tabular}{ccccc}
    \hline
    label & stable trivial   & $\#$ non-trivial  & $\#$ stable non-trivial  & structural \\
    & configuration &  configurations & configurations & response\\
    \hline
    $\mathsf{O}'$  & no & 0 & 0 & only unstable confs \\
    $\mathsf{O}$  & yes & 0 & 0 & \emph{monostable} \\
    $\mathsf{A}$  & yes & 2 & 1 & \emph{bistable} \\
    $\mathsf{B}$  & yes & 8 & 3 & \emph{tetrastable} \\
    $\mathsf{C}$  & yes & 3 & 0 & \emph{monostable} \\
    $\mathsf{D}$  & no & 3 & 1 & \emph{monostable}  \\
    $\mathsf{E}$  & no & 1 & 1 & \emph{monostable}  \\
    $\mathsf{F}$  & no & 1 & 0 & only unstable confs   \\
    $\mathsf{G}$ & no & 5 & 2 & \emph{bistable} \\
    \hline
    \end{tabular}
    \end{center}
\end{table}

It can be observed that the number of stable configurations can be reduced by increasing the angle $\beta_0$, for example the \emph{tetrastability} region disappears, while the region without stable equilibrium solutions expands. 
Moreover, roughly speaking, the monostable region $\mathsf{O}$ is  mirrored around the $\mathcal{H}$--axis, when $\beta_0$ moves from the range ($0, 90^\circ$) to the range  ($90^\circ, 180^\circ$).
It is also interesting to note that the system may turn from \emph{monostable} to  \emph{tetrastable} without displaying an intermediate \emph{bistable} behaviour, as in the case reported in Fig. \ref{fig:bifurcation_regions}($a$) by  decreasing values of $\mathcal{V}$ at $\mathcal{H}=0$.

\begin{figure}[!h]
	\centering
	\includegraphics[width=\textwidth]{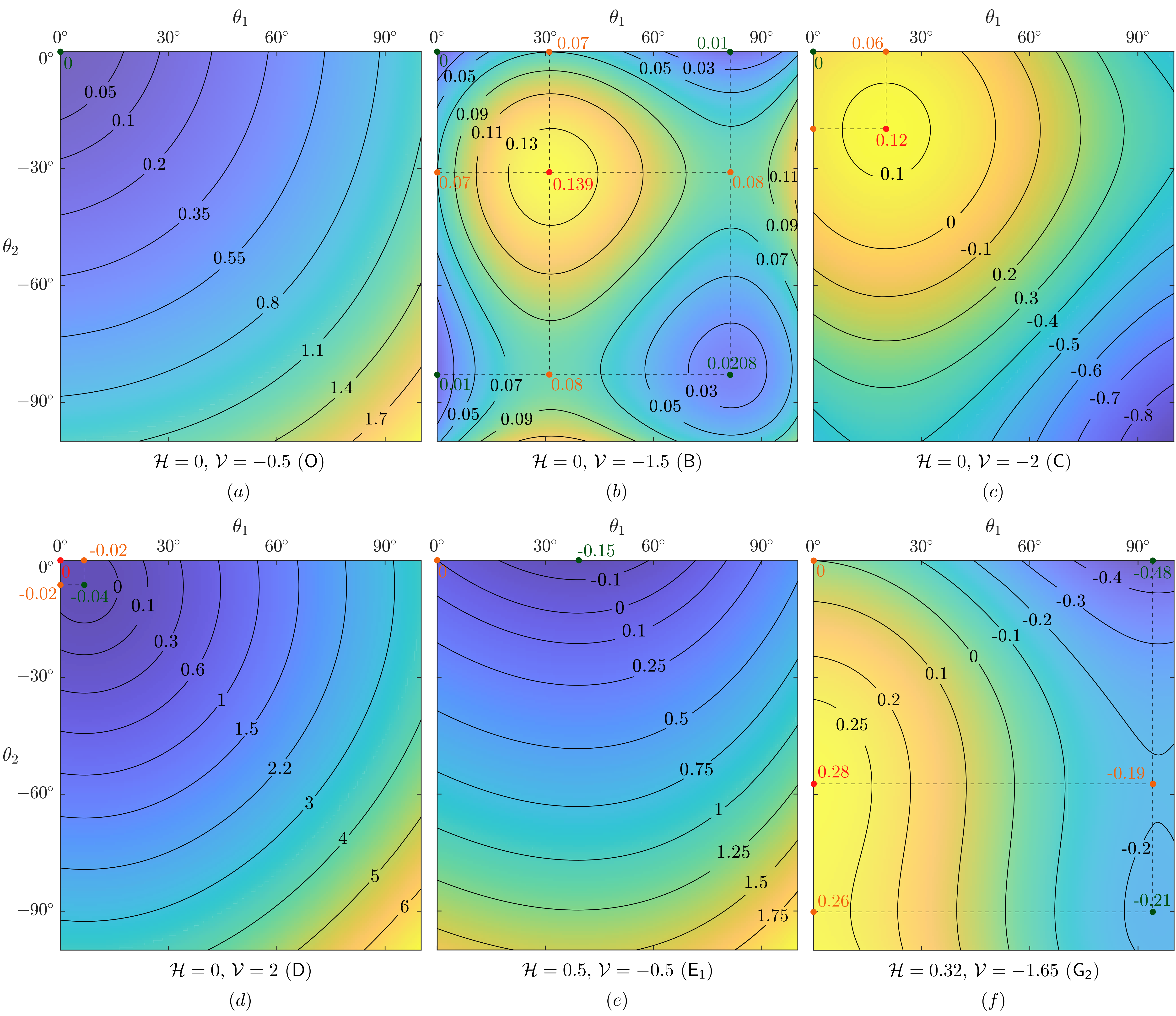}
	\caption{
	Contourplots of the dimensionless total potential energy $\Pi/K_2$, eq. \eqref{TOTpot_eng}, on the  difference angles plane $\theta_1$--$\theta_2$ for a system defined by the parameters set ($f$) of  Table \ref{tab:cases} and for six different pairs of loads $\mathcal{H}-\mathcal{V}$.  By varying the applied loads, the number of  total potential energy wells changes, defining a ($a$, $c$, $d$, $e$) \emph{monostable}, ($f$) \emph{bistable}, and ($b$) \emph{tetrastable} system. Green, red, and orange circles respectively define local minima, local maxima, and saddle points, and therefore the first  correspond to stable equilibrium configurations, while the second and third to unstable ones. 
	}
	\label{potential_wells}
\end{figure}

To further appreciate the generation of more than one stable configurations, the contourplots of the total potential energy $\Pi$, eq. \eqref{TOTpot_eng}, on the  difference angles  plane $\theta_1$--$\theta_2$ are reported in Fig. \ref{potential_wells} for the parameters set ($f$) of Table \ref{tab:cases}.
The contourplots for six different pairs of $\mathcal{H}-\mathcal{V}$ loads show how the number of  total potential energy wells changes, defining ($a$, $c$, $d$, $e$)  \emph{monostable}, ($f$) \emph{bistable}, or ($b$) \emph{tetrastable} systems.

\begin{figure}[!h]
	\centering
	\includegraphics[width=\textwidth]{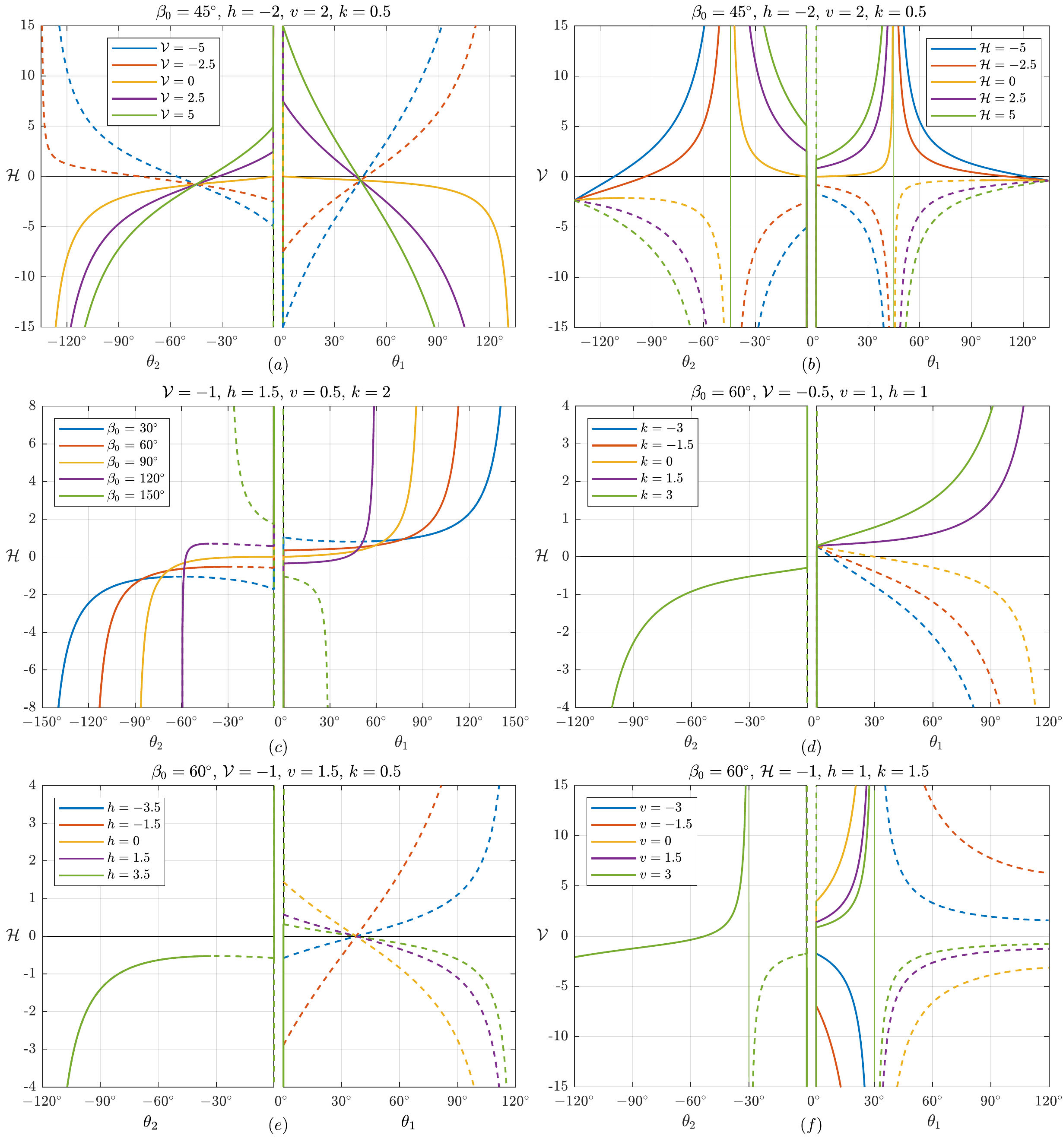}
	\caption{Equilibrium configurations in terms of difference angles $\theta_1$ and $\theta_2$ versus the load  $\mathcal{H}$ or $\mathcal{V}$ at fixed value of the remaining parameters ($\mathcal{V}$ or $\mathcal{H}$, $h$, $v$, $k$,  and $\beta_0$) as specified in each panel. Stable and unstable configurations are displayed as continuous and dashed curves, respectively.}
	\label{fig:eq_paths}
\end{figure}

Equilibrium configurations expressed as 
horizontal $\mathcal{H}$ and 
the vertical  $\mathcal{V}$ loads, functions 
of the 
difference angles $\theta_1$ and $\theta_2$ are reported in 
Fig. \ref{fig:eq_paths}, at fixed values of the remaining parameters.
Stable and unstable configurations are distinguished through continuous and  dashed lines. The reported curves show how significantly the different parameters affect the critical loads, the post-buckling response of the structure (by turning the incremental stiffness from positive to negative and by realizing force-reversal conditions), and the stability. More specifically,
\vspace{-2mm}
\begin{itemize}
    \item bifurcation of  both layers occurs at the unloaded state ($\mathcal{H}=\mathcal{V}=0$), Fig.  \ref{fig:eq_paths}($a$, $b$);
    \item at constant  vertical load $\mathcal{V}$, the bifurcation load $\mathcal{H}_{cr}$ and the post-buckling behaviour can be tuned also by varying the  configuration angle $\beta_0$, Fig.  \ref{fig:eq_paths}($c$);
    \item 
    when the deformed $j$-th layer assumes the rectangular shape, condition occurring for $\theta_j=-(-1)^j(\pi/2-\beta_0)$, the following situations occur:
    \begin{itemize}
    \item the equilibrium is  independent of the vertical loads ($\mathcal{V}$ and $v$), Fig.  \ref{fig:eq_paths}($a$),   when 
  \begin{equation}\label{no_V_influence}
    	\begin{array}{lll}
    	    \mathcal{H} = \left(\dfrac{\pi }{2}-\beta _0\right) \dfrac{k }{(h+1)},& \forall\,\, \mathcal{V}\,\, \mbox{and}\,\, v & \mbox{ for } \theta_1 = \dfrac{\pi}{2}-\beta_0, \\[5mm]
    	    \mathcal{H} = -\left(\dfrac{\pi }{2}-\beta _0\right),& \forall\,\, \mathcal{V}\,\, \mbox{and}\,\, v & \mbox{ for } \theta_2 = -\dfrac{\pi}{2}+\beta_0;
    	\end{array}
\end{equation}
\item when the horizontal force $\mathcal{H}$ does not satisfy eq. (\ref{no_V_influence}), the  rectangular configuration is attained only at an infinite value of the vertical load $\mathcal{V}$, Fig. \ref{fig:eq_paths}($b$, $f$), representing a \emph{locking condition} for the system;
 \end{itemize}
\item  when $h<0$, a stable (an unstable) post-critical  behaviour can be associated to negative (positive) slope in the equilibrium path   $\mathcal{H}-\theta_j$,  
Fig. \ref{fig:eq_paths}($a$, $e$); 
\item when $\mathcal{H}=0$, the equilibrium angle $\theta_1$ is independent of $h$, Fig. \ref{fig:eq_paths}$(e)$;
        \item  the stiffness and loading ratios $k$, $h$ and $v$ influence the bifurcation load and  the post-critical behaviour of the lower layer only, Fig. \ref{fig:eq_paths}($d$, $e$, $f$);
\item  at null  horizontal load  ($\mathcal{H}=0$), both layers bifurcate simultaneously when the  vertical load $\mathcal{V}$ is greater than zero for $\beta_0\in(0,\pi/2)$, Fig. \ref{fig:eq_paths}($b$), or is smaller than zero for $\beta_0\in(\pi/2,\pi)$.
\end{itemize}

\paragraph{Symmetric response.}
When the stiffness ratio $k$ and the loading ratios $h$ and $v$ satisfy the following condition 
\begin{equation}\label{symcon}
k=v+1=h+1,
\end{equation}
the post-critical response defined by eq. (\ref{static_eqn_theta}) reduces to
\begin{equation}\label{equilibrium_beta}
    -(-1)^j\mathcal{H} \sin \left(\beta _0 + |\theta_j| \right) + \mathcal{V} \cos \left(\beta _0 + |\theta_j| \right) = |\theta_j|,
\end{equation}
which implies the following
 symmetry property
\begin{equation}\label{symsym}
\theta_1(\mathcal{H} ,\mathcal{V} )=-\theta_2(-\mathcal{H} ,\mathcal{V} ).
\end{equation}

It is finally noted that, due the positiveness of $k$, the condition (\ref{symcon}), defining the symmetric response (\ref{symsym}), may be realized only when
\begin{equation}
h>-1 \quad \mbox{and} \quad v>-1,
\end{equation}
which includes the particular case of rotational springs of equal   stiffness and absence of forces acting on the top of the lower layer,
\begin{equation}\label{symmetry_condition}
		h=0, \qquad v=0, \qquad k=1. 
\end{equation}

\subsection{Stability of the equilibrium}\label{stability}

According to the Dirichlet criterion, stability corresponds to the positive definiteness of the Hessian (symmetric) matrix $\mathbb{H}$ of the total potential energy $\Pi$, whose components are given by
\begin{equation}\label{symmetry_matrix}
	\begin{array}{cc}
	\mathbb{H}_{ij}= \displaystyle \sum_{q=1}^{2}\sum_{r=1}^{2}
	\left( \dfrac{\partial^2 \Pi}{\partial\theta_q \partial\theta_r}\dfrac{\partial\theta_q}{\partial\varphi_j}
	\dfrac{\partial\theta_r}{\partial\varphi_i}
	+\dfrac{\partial \Pi}{\partial\theta_q}\dfrac{\partial^2\theta_q}{\partial\varphi_j \partial\varphi_i}\right).
	\end{array}
\end{equation}
Considering the total potential energy  $\Pi$ expressed by eq.  \eqref{TOTpot_eng} and the property in eq. \eqref{no_coupling}, the following conditions holds
\begin{equation}
    \begin{array}{cc}
\dfrac{\partial^2\Pi}{\partial\theta_i \partial\theta_j} = 0,
\qquad i \neq j.
    \end{array}
\end{equation}
Therefore, the Hessian matrix is diagonal ($\mathbb{H}_{12}=\mathbb{H}_{21}=0$) and its eigenvalues $\mu_j$ coincide with the corresponding diagonal terms, 
\begin{equation}
\mu_j = \mathbb{H}_{jj} = \dfrac{\partial^2\Pi}{\partial\theta_j^2} \left(\dfrac{\partial\theta_j}{\partial\varphi_j}\right)^2 + \dfrac{\partial\Pi}{\partial\theta_j}\dfrac{\partial^2\theta_j}{\partial\varphi_j^2},
\end{equation}
where
\begin{equation}\label{deriva}
         \dfrac{\partial^2\theta_j}{\partial\varphi_j^2} =   
        -(-1)^j \dfrac{2 \; \psi \left[\dfrac{\sin ^2\beta _0+\psi ^2 \tan ^4\varphi_j}{\cos ^2\varphi_j}  + 2 \tan^2 \varphi_j \left(1-\left(\cos \beta _0 -\psi  \tan ^2\varphi_j \right)^2\right) \right]}{\cos ^2\varphi_j \sqrt{\left[1-\left(\cos \beta _0 -\psi  \tan ^2\varphi_j \right)^2\right]^{3}}}  .
\end{equation}

It is interesting to note that each eigenvalue only depends on the respective difference angle, namely
\begin{equation}
\mu_j=\mu_j(\theta_j),
\end{equation}
and therefore, as for the  equilibrium, the stability analysis of the equilibrium configuration is decoupled for the two layers. Considering that  the conditions of stability, called \lq S$_j$', and instability, called \lq  U$_j$', of  the equilibrium configuration $\theta_j$ for the $j$-th layer are given by
\begin{equation}\label{assstab}
\left.\begin{array}{cc}
			 \text{S}_j \\[3mm]
    		 \text{U}_j\\
    		\end{array}
    		\right\} :	\sgn[\mu_j(\theta_j)] \left\{\begin{array}{cc}
			 > 0, \qquad & \text{Stable} \\[3mm]
    		 < 0, \qquad & \text{Unstable} \\
    		\end{array}
    		\right\}  \mbox{ configuration for the $j$-th  layer},
\end{equation}
it follows that
\emph{the stability of the equilibrium configuration for the two-layer unit structure  is provided by the simultaneous \lq individual' stability of the  configuration  assumed by each layer, corresponding to $\text{S}_1 \cap\text{S}_2$.}

Considering all of the above, the stability of the trivial and non-trivial configurations is addressed separately.

\paragraph{Trivial configuration.} When $\theta_j = 0$, the first and second derivatives of the difference angle $\theta_j$ with respect to the misalignment angle $\varphi_j$,
eqs. \eqref{no_coupling} and \eqref{deriva}, reduce to

\begin{equation}
        \dfrac{\partial\theta_j}{\partial\varphi_j}= 0, \qquad
         \dfrac{\partial^2\theta_j}{\partial\varphi_j^2} = -(-1)^j \dfrac{2 \; \psi }{\sin \beta_0},
\end{equation}
and the eigenvalues become
\begin{equation}\label{trivial_eigen}
        \mu_j = \dfrac{\partial\Pi}{\partial\theta_j}\dfrac{\partial^2\theta_j}{\partial\varphi_j^2}.
\end{equation}
Therefore, from eq. \eqref{trivial_eigen} the stability conditions (\ref{assstab}) for the trivial equilibrium path  read as 
\begin{equation}\label{trivial_stablity}
\begin{array}{ll}
\left.\begin{array}{cc}
			 \text{S}_1 \\[3mm]
    		 \text{U}_1 \\
    		\end{array}
    		\right\} :	(h+1) \mathcal{H} \sin \beta_0+(v+1) \mathcal{V} \cos \beta_0  \left\{\begin{array}{cc}
			 < 0, \quad & \text{Stable} \\[3mm]
    		 > 0, \quad & \text{Unstable} \\
    		\end{array}
    		\right\} \begin{array}{ll} \mbox{trivial configuration}\\[0mm] \mbox{ of the lower layer},
    		\end{array}
    		\\[8mm]
    	\left.\begin{array}{cc}
			 \text{S}_2 \\[3mm]
    		 \text{U}_2 \\
    		\end{array}
    		\right\} :
    		\mathcal{H} \sin \beta _0 - \mathcal{V} \cos \beta _0  \left\{\begin{array}{cc}
    		> 0, \quad & \text{Stable} \\[3mm]
    		< 0, \quad & \text{Unstable}
		\end{array}\right\}\mbox{ trivial configuration of the upper layer},
	\end{array}
\end{equation}
which is used to define  the stable or unstable character of the trivial equilibrium configuration $\theta_j=0$ displayed in Figs. \ref{introcurves}, \ref{fig:eq_surf}, \ref{fig:bifurcation_regions}, \ref{fig:eq_paths}.
As a consequence,  the trivial configuration of the two-layer unit structure is stable when 
\begin{equation}\label{s1s2gen}
\text{S}_1 \cap\text{S}_2 :    (h+1) \mathcal{H} \sin \beta_0+(v+1) \mathcal{V} \cos \beta_0<0,
    \quad
    \mathcal{H} \sin \beta _0 - \mathcal{V} \cos \beta _0>0.
\end{equation}

Four regions of loading combinations $\mathcal{H}$--$\mathcal{V}$ can be distinguished according to the stability of the trivial configuration assumed by the two layers, 
and corresponding to the following situations: (i.) stable undeformed configuration for both layers ($\text{S}_1 \cap\text{S}_2$);  (ii.) unstable undeformed configuration for both layers ($\text{U}_1 \cap\text{U}_2$); 
(iii.) unstable undeformed configuration for the lower layer and stable for the upper one ($\text{U}_1 \cap\text{S}_2$); 
(iv.) unstable undeformed configuration for the upper layer and stable for the lower ($\text{S}_1 \cap\text{U}_2$).

Finally, it is interesting to note that  the  stability condition (\ref{s1s2gen}) reduces in the case of rectangular undeformed layers  ($\beta_0=\pi/2$) to 
    \begin{equation}
\text{S}_1 \cap\text{S}_2 :  \mathcal{H} >0, \quad  h<-1,
    \quad \forall\,\,\mathcal{V}.
\end{equation}

\paragraph{Non-trivial configuration.} When $\theta_j \neq 0 $, the equilibrium condition reduces to
\begin{equation}
    \begin{array}{lll}
         \dfrac{\partial\Pi}{\partial\theta_j} = 0,
    \end{array}
\end{equation}
therefore the $j$-th eigenvalue simplifies to
\begin{equation}\label{non_trivial_eigen}
    \mu_j = \dfrac{\partial^2\Pi }{\partial\theta_j^2} \left(\dfrac{\partial\theta_j}{\partial\varphi_j}\right)^2,
\end{equation}
and its sign coincides with that of the second derivative of the total potential energy, 
\begin{equation}\label{non_trivial_eigen_sign}
    \sgn\left[\mu_j\right] = \sgn\left[\dfrac{\partial^2\Pi }{\partial\theta_j^2}\right].
\end{equation}
The sign of the second derivative of the total potential energy with respect to $\theta_j$ defines the stable or unstable character of the non-trivial equilibrium configuration for the difference angle $\theta_j$, as displayed in Figs. \ref{introcurves}, \ref{fig:eq_surf}, \ref{fig:bifurcation_regions}, \ref{fig:eq_paths}. Due to the high nonlinearities, the second derivative of the total potential energy, evaluated at the non-trivial equilibrium configuration, is given by a complicated expression and therefore is omitted. However, a first-order expansion in the difference angle amplitude allows to evaluate this quantity through the following  expression
\begin{equation}\label{eq_bif_cond_expanded}
	\begin{array}{lll}
		\left.\dfrac{\partial^2\Pi }{\partial\theta_j^2} \,\,\right|_{\mathcal{H}=\mathcal{H}(\theta_j,\mathcal{V})}=
		\left(k^{2-j}+\dfrac{(v+1)^{2-j} \,  \mathcal{V}}{\sin \beta _0}\right)\left(1-\dfrac{\left| \theta_j \right|}{\tan \beta_0}\right),
	\end{array}
\end{equation}
and the stability conditions (\ref{assstab}) reduce for small difference angles ($|\theta_j|\gg\theta_j^2$) to 
\begin{equation}\label{non-trivial_stablity}
\begin{array}{ll}
\left.\begin{array}{cc}
			 \text{S}_1 \\[3mm]
    		 \text{U}_1 \\
    		\end{array}
    		\right\} :	k+\dfrac{(v+1)\, \mathcal{V} }{\sin \beta _0} \left\{\begin{array}{cc}
			 > 0, \qquad & \text{Stable} \\[3mm]
    		 < 0, \qquad & \text{Unstable} \\
    		\end{array}
    		\right\}  \mbox{non-trivial configuration for the lower layer,}\\[8mm]
\left.\begin{array}{cc}
			 \text{S}_2 \\[3mm]
    		 \text{U}_2 \\
    		\end{array}
    		\right\} :    			1+\dfrac{\mathcal{V} }{\sin \beta _0}
    		 \left\{\begin{array}{cc}
    		> 0, \qquad & \text{Stable} \\[3mm]
    		< 0, \qquad & \text{Unstable}
		\end{array}\right\}\mbox{non-trivial configuration for the upper layer,}
	\end{array}
\end{equation}
showing that just after the bifurcation, the stability of the non-trivial path is affected also by the stiffness ratio $k$ and the load ratios $h$ and $v$, in addition to the angle $\beta_0$.

\section{Conclusions}

The nonlinear quasi-static mechanical behaviour of a  structural element, to be used as a unit structure for metamaterial design, has been analyzed. 
The structural element exhibits a complex bifurcation landscape, with multiple (stable and unstable)  equilibrium configurations as related to the presence of an element  susceptible to buckling at vanishing load under tension or compression. 
This element introduces a mechanical equivalence with a unilateral constraint, thus strongly conditioning dynamics, which will be analyzed elsewhere. 

\vspace*{5mm} \noindent {\sl Authors' Contributions.} The presented structure was designed and developed by all authors.
They also cooperated in the development of calculations and in the writing of the paper. 
In addition, NH created all the codes for symbolic manipulation and numerical results. He also prepared all the figures.

\vspace*{5mm} \noindent {\sl Acknowledgments.} NH and FDC gratefully acknowledge the financial support from the European Union’s Horizon 2020 research and innovation programme under the Marie Sklodowska-Curie grant agreement ‘INSPIRE - Innovative ground interface concepts for structure protection’ PITN-GA-2019-813424-INSPIRE.
DB gratefully acknowledges financial support from the ERC advanced grant
ERC-2021-AdG-101052956-BEYOND. Support 
from the Italian Ministry of Education,
University and Research (MIUR) in the frame of the
‘Departments of Excellence’ grant L. 232/2016 is acknowledged.  This work has been developed under the auspices of INDAM-GNFM.
 \vspace*{10mm}

\bibliography{bibliography_plain}
\bibliographystyle{ieeetr}

\end{document}